\documentclass[a4paper]{article}
\usepackage{authblk}
\usepackage[utf8]{inputenc}
\usepackage{graphicx}
\usepackage[text={17cm,26cm},centering]{geometry}
\usepackage{amsmath}
\usepackage[final]{pdfpages}
\usepackage{import}
\usepackage{bibunits}
\graphicspath{{figs/}}

\newcommand{\Ee}{E_\mathrm{e}}
\newcommand{\Eg}{E_\gamma}
\newcommand{\yg}{y_\gamma}
\newcommand{\xg}{x_\gamma}
\newcommand{\Lec}{L_\mathrm{ec}}
\newcommand{\yc}{y_\mathrm{c}}
\usepackage{commath}

\title{Spontaneous surface reserve formation in wicked membranes bestow extreme stretchability}

\author[1]{Paul Grandgeorge}
\author[2]{Natacha Krins}
\author[1,3]{Aurélie Hourlier-Fargette}
\author[2]{Christel Laberty}
\author[1]{Sébastien Neukirch}
\author[1,4]{Arnaud Antkowiak}
\affil[1]{Sorbonne Universités, UPMC Univ Paris 06, CNRS, UMR 7190 Institut Jean Le Rond d'Alembert}
\affil[2]{Sorbonne Universités, UPMC Univ Paris 06, CNRS, UMR 7574 Laboratoire de chimie de la matière condensée de Paris}
\affil[3]{Département de Physique, École Normale Supérieure, CNRS, PSL Research University, F-75005 Paris, France}
\affil[4]{Surface du Verre et Interfaces, UMR 125 CNRS/Saint-Gobain, F-93303 Auberviliers, France}

\begin{document}
\begin{bibunit}[naturemag]
\maketitle

\textbf{
Soft stretchable materials are key for arising technologies such as stretchable electronics \cite{Rogers2010} or batteries \cite{Liu2016}, smart textiles \cite{Hu2012}, biomedical devices \cite{Minev2015}, tissue engineering and soft robotics \cite{Shepherd2011,Lazarus2015}.
Recent attempts to design such materials, via e.g. micro-patterning of wavy fibres \cite{Ahn2009} on soft substrates, polymer engineering at the molecular level \cite{sun2012highly} or even kirigami techniques \cite{rafsanjani2017buckling}, provide appealing prospects but suffer drawbacks impacting the material viability: complexity of manufacturing, fatigue or failure upon cycling, restricted range of materials or biological incompatibility.
Here, we report a universal strategy to design highly stretchable, self-assembling and fatigue-resistant synthetic fabrics. Our approach finds its inspiration in the mechanics of living animal cells that routinely encounter and cope with extreme deformations, e.g. with the engulfment of large intruders by macrophages \cite{Lam2009}, squeezing and stretching of immune cells in tiny capillaries \cite{Guillou2016} or shrinking/swelling of neurons upon osmotic stimuli \cite{Morris2001}. All these large instant deformations are actually mediated and buffered by membrane reserves available in the form of microvilli, membrane folds or endomembrane that can be recruited on demand \cite{Guillou2016, Raucher1999}.
We synthetically mimicked this behavior by creating nanofibrous liquid-infused tissues spontaneously forming surface reserves whose unfolding fuels any imposed shape change.
Our process, relying only on geometry, elasticity and capillarity, allows to endow virtually any material with high stretchability and reversibility, making it straightforward to implement additional mechanical, electrical or chemical functions.
We illustrate this with proof-of-concept activable capillary muscles, adaptable slippery liquid infused porous surfaces and stretchable basic printed electronic circuits.
}

\noindent Geometry and elasticity of thin objects are intimately linked, so that metallic wires with identical diameter but with different shapes (e.g. straight or curly as a spring) will contrast markedly in their mechanical response.
Nature abounds in such examples where mechanical behavior is entangled with geometry.
Leaf geometrical curvature, for example, is critical for some carnivorous plants' prey-trapping ability. For example, the particular shape adopted by the Venus flytrap's leaf brings it on the verge of an elastic instability, requiring from an insect only a minute stroke to snap and rush the vegetal jaws -- thereby providing the plant with one of the fastest non-muscular movements \cite{Forterre2005}.
\begin{figure}[h!]
\begin{center}
\includegraphics{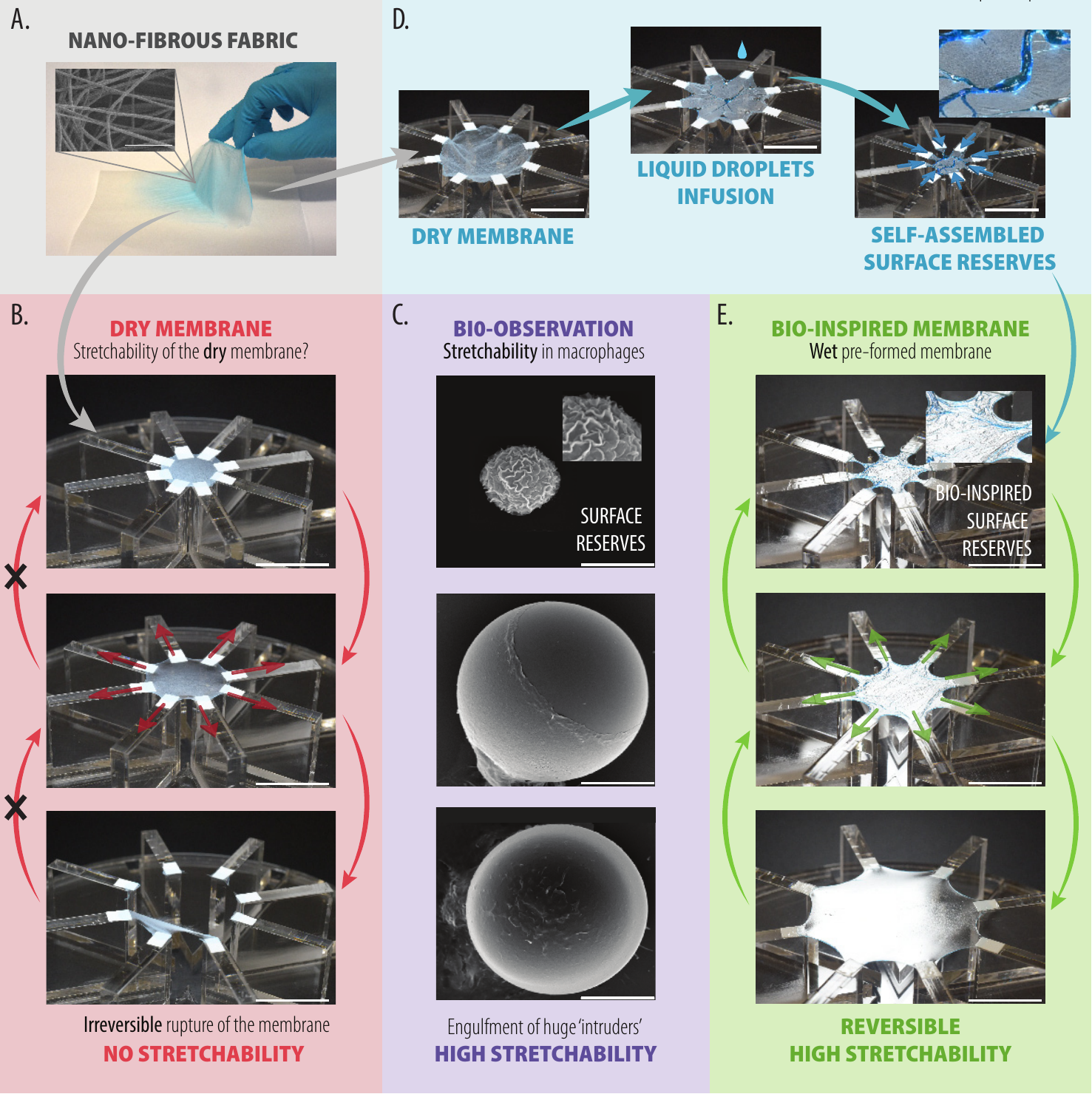}
\end{center}
\caption{\textbf{Designing ultra-stretchable membranes.} \textbf{A.} A thin fibrous membrane (here: electrospun PVDF-HFP membrane dyed blue with a coloring agent for visualization purposes) and a scanning electron microscope micrograph of this membrane are presented (scale bar: 50 $\mu$m). The typical diameter of the fibers composing this few microns thick membrane is 500 nm. \textbf{B.} At its native (dry) state, the membrane is attached to 8 translational supports slowly separated from each other. As the membrane is not intrinsically stretchable, it rapidly irreversibly tears off at around 30\% area extension. Scale bar: 3 cm. \textbf{C.} SEM micrographs of a J774 macrophage (Courtesy of Prof. Heinrich, UC Davis \cite{Lam2009}). The first snapshot presents the macrophage at rest (cell size is not representative due to shrinkage during fixation). The two following pictures show the macrophage engulfing an antibody coated (FC$\gamma$-opsonized) 30 $\mu$m diameter bead. Throughout this ambitious engulfment, the initially corrugated cell membrane smooths out, thus recruiting membrane surface, necessary for its up to 5-fold surface area expansion. Scale bar: 10 $\mu$m. \textbf{D.} Based on this biological observation, our membrane is now attached to the 8 translational supports (set in a wide position) and wicked by a wetting liquid (here: v100 silicone oil). When the supports are brought closer, the membrane does not sag as one would expect; it spontaneously folds inside the liquid veins due to the surface tension they develop. Scale bar: 3 cm. \textbf{E.} Once the whole membrane is wicked by the liquid, its compressed state spontaneously adopts a wrinkled and folded surface, similar to that of the resting J774 membrane. Just like for the J774 macrophage, our self-corrugated wicked membrane smooths out when stretched; wrinkles and folds act as membrane reserves that can fuel up to a 50-fold surface area expansion. Moreover, as the process relies only on elasticity and capillarity, it is reversible and repeatable (wrinkles and folds continue self-assembling upon subsequent compressions). Scale bar: 3 cm. }
\label{fig:stretchable_fabric}
\end{figure}
Geometry is also behind the curious mechanical behavior of the capture silk spun by ecribellate spiders: whether stretched or compressed, this fibre remains straight while seemingly adjusting its length, as if telescopic.
Actually the pulling force of surface tension allows to coil, spool and pack excess fibre within the glue droplets decorating the thread.
These so formed fibre reserves can then be recruited on demand -- and confer the thread an apparent extreme strechability of +10,000\% \cite{elettro2016drop}.
Another example can be found inside our own body with cells which display a particular ability to cope with stretch. Macrophages extending their surface area by a factor 5 to engulf large microbes or cellular debris \cite{Lam2009} (see Fig.~\ref{fig:stretchable_fabric}), patrolling T-lymphocytes stretching by 40\% to squeeze into the microvasculature \cite{Guillou2016}, hundreds of $\mu$m sized neuronal projections extruded from 10 $\mu$m wide neurons \cite{Raucher1999, Dai1995} or osmotic swelling of fibroblast leading to 70\% increase in area \cite{Groulx2006} are a few out of many examples of the extreme mechanical sollicitations encountered by living animal cells on a routine basis. This resilience is all the more spectacular that the lytic stretching level at which the plasma membrane ruptures is about 4\% \cite{Raucher1999, Nichol1996}. Why do these cells then just not burst under stress?
Actually cells have evolved a strategy consisting in storing excess membrane in the form of folds and microvilli \cite{Erickson1976,Majstoravich2004} which can be recruited and deployed on demand. Interestingly, these geometrical corrugations enabling stretching do not fluff the membrane. Cellular tension is indeed preserved thanks to the pulling action of the underlying cortical actin layer \cite{Salbreux2012}.

\noindent Learning from these examples, we here transpose Nature's blueprints by making use of self-assembled membrane reserves to endow synthetic fabrics with high stretchability. Figure~\ref{fig:stretchable_fabric} illustrates the key steps to design such an extensible tissue. We first manufacture, with a conventional spinning technique, a light and free-standing non-woven fabric. Without further treatment, the so-formed fibrous membrane would show early signs of damage above a few percents of extension, and would definitely rupture at a 30\% area extension.
In order to mimick the pulling action of the cortical actin layer, we infuse the polymeric mat with a wetting liquid so as to bestow the resulting wick with surface tension.
Instantaneously, the membrane self-tenses while seemingly adjusting its surface, storing any excess membrane into folds: the membrane reserves.
Once formed, these geometrical ruffles and furrows can be unfolded at will, fueling any imposed shape change to the membrane, see Fig.~\ref{fig:stretchable_fabric}. Remarkably, the process is entirely reversible as the membrane reserves self-assemble instantaneously upon contraction.
\begin{figure}[htb]
\begin{center}
\includegraphics[width=16cm]{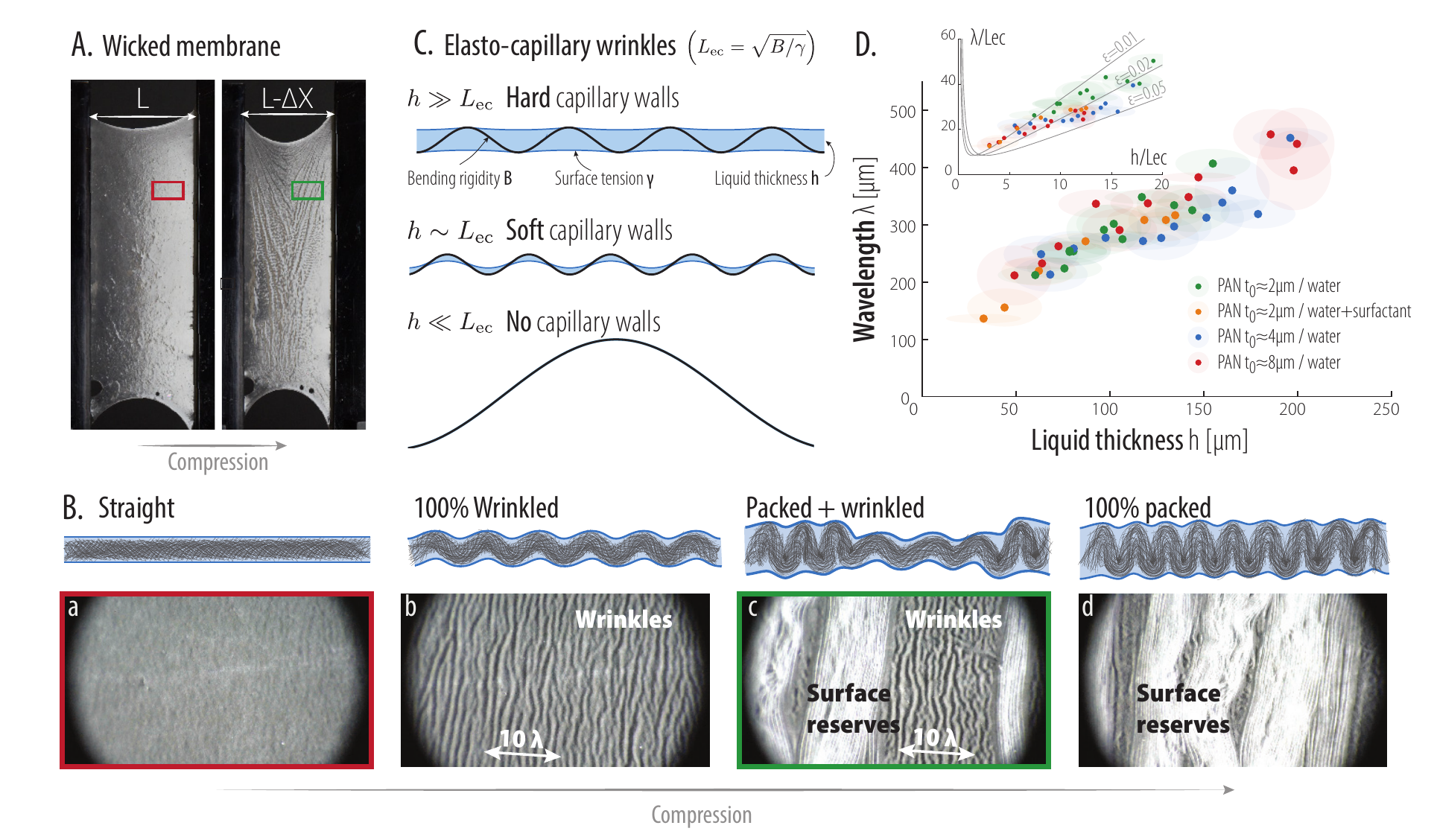}
\end{center}
\caption{\textbf{
Mechanics of the wicked membrane: capillary-driven wrinkling and packing.
} \textbf{A.} Electro-spun PAN membrane wicked with water and attached to two rigid mobile edges. The left and right photo respectively show the membrane at its extended and slightly compressed state. Upon this small compression, the wicked membrane immediately exhibits a clear wrinkling pattern ($L$=4 cm). \textbf{B.} Closer view of the membrane (red and green rectangles in {A.}) throughout a complete uni-axial compression. At the extended state, the wicked membrane is smooth (a). A small compression engender a wrinkled surface (of wavelength $\lambda$) which rapidly displays a two phase texture as compression is pursued (b and c respectively). One phase corresponds to a wrinkled texture (same wavelength $\lambda$) and the other one a closely packed accordion-like state which gradually expands throughout compression. The whole membrane is closely packed inside the liquid film at the end of the compression (d). This accordion-like phase houses the surface reserves which will be recruited upon a subsequent extension of the membrane. \mbox{(scale: 10$\lambda$=3.1 mm)} \textbf{C.} Physical interpretation of the early wrinkling of the membrane inside the liquid film. Here, the membrane is described as an elastic beam of bending stiffness per unit depth $B$ inside a liquid film of initial thickness $h$ and liquid-air interface energy $\gamma$. Considering the beam buckles in a sinusoidal shape and the liquid interface adopts a circular shape, at a given compression~$\epsilon$, three regimes emerge, depending on the liquid film thickness. \textbf{D.} Experimental wavelength $\lambda$ of the wrinkles observed at early an compression stage of the membrane as a function of the liquid film thickness $h$ for different membrane thicknesses $t_0$ and infusing liquids. The inset provides the data normalized by the elasto-capillary length $L_\mathrm{ec}$. The solid gray lines correspond to an energy minimization analysis for compressions of $\epsilon=1,\,2$ and $5$\% (see Supplementary Material). }
\label{fig:membrane_reserve_mechanics}
\end{figure}

To shed light over the mechanics of membrane reserve self-assembly, we now investigate the behavior of the wicked membrane at the microstructural level, i.e. at lengthscales of the order of the liquid film thickness.
Figure~\ref{fig:membrane_reserve_mechanics} reports typical global and close-up views of a membrane undergoing compression. It is readily seen that whenever compression starts, the initially flat membrane (Fig.~\ref{fig:membrane_reserve_mechanics}A left and Fig.~\ref{fig:membrane_reserve_mechanics}Ba) is rapidly textured with a wrinkling pattern exhibiting a clear wavelength $\lambda$ (Fig.~\ref{fig:membrane_reserve_mechanics}A right and Fig.~\ref{fig:membrane_reserve_mechanics}Bb).
Wrinkling is a trademark of thin elastic sheets, and develops spontaneously in a variety of contexts: pinched skin, shriveling fruits \cite{Cerda2003}, brain sulci \cite{Tallinen2016}, hanging curtains \cite{Vandeparre2011} or more generally thin sheets under tension \cite{Davidovitch2011}.
This elastic instability occurs whenever a compressed slender structure is bound to a substrate resisting deformation.
The emerging wavelength $\lambda$ of this particular form of buckling therefore results from a trade-off between the deformation of the membrane and that of the substrate in order to minimize global energy.
Here, the substrate role is played by the liquid film interfaces, which can be seen as soft capillary walls restraining the deformation of the membrane.
Interestingly, experiments reveal that the wrinkles wavelength $\lambda$ is neither particularly sensitive to the value of interfacial tension $\gamma$ nor the fibrous membrane thickness $t_0$, but scales linearly with the liquid film thickness $h$, here measured by means of colorimetry (see Fig.~\ref{fig:membrane_reserve_mechanics}D and Materials \& Methods).
In order to further grasp the physics of wrinkle formation, we develop a simple model where a periodic sinusoidal membrane of bending stiffness per unit depth $B$ exhibiting a wavelength $\lambda$ interacts with a liquid film exposing two free surfaces of surface tension $\gamma$, see Fig.~\ref{fig:membrane_reserve_mechanics}B.
\begin{figure}[htb]
\begin{center}
\includegraphics[width=16cm]{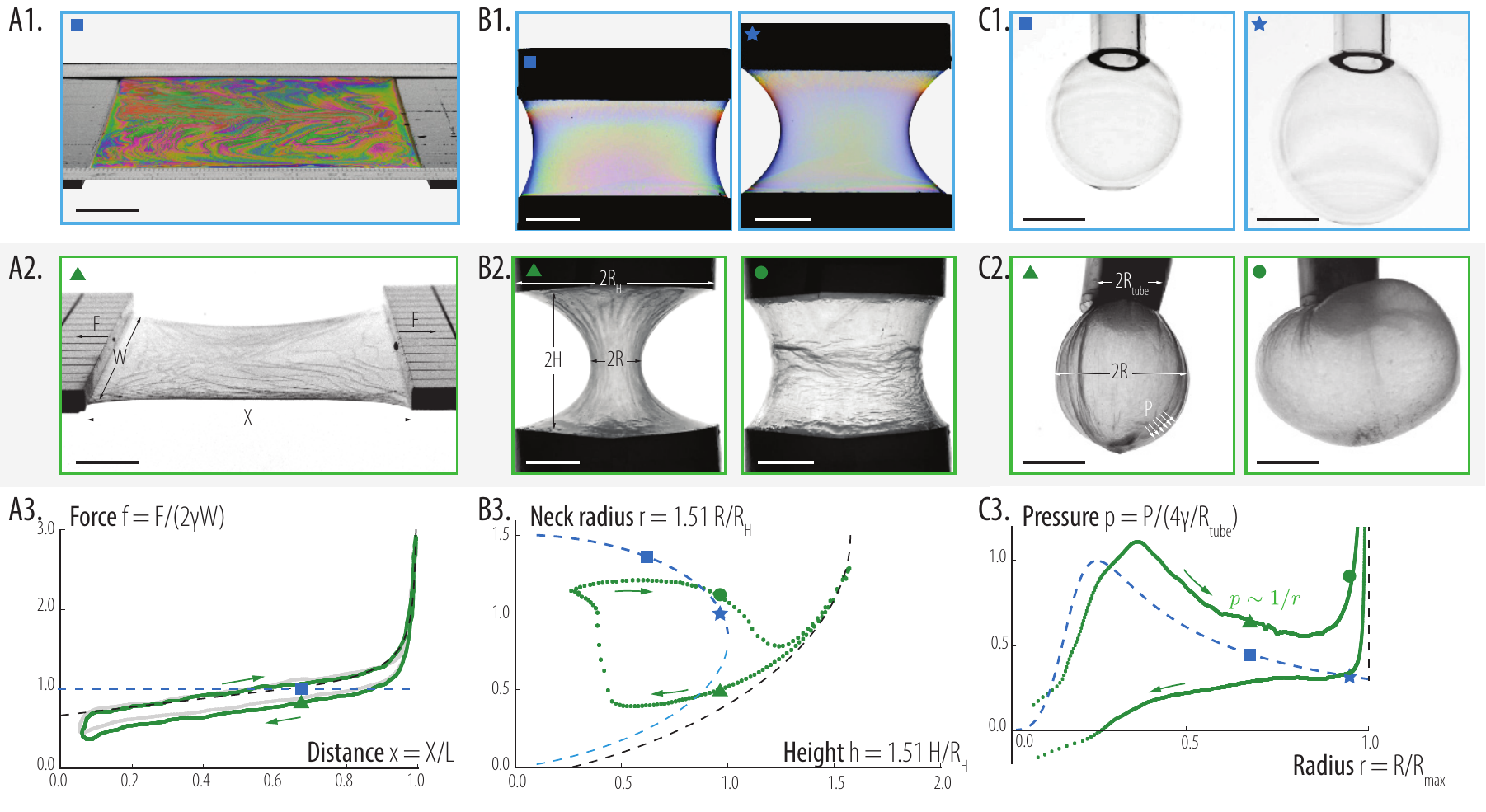}
\end{center}
\caption{\textbf{Forms and forces for capillary-folded wicked membranes}. \textbf{A1.} Soapy liquid film on a frame. \textbf{A2.} Wicked membrane attached to two mobile supports. The membrane has a width $W$ and its ends are separated by a distance $X$. \textbf{A3.} Force versus displacement diagram. The green curve corresponds to the force measurements during the first compression/extension cycle of the wicked membrane whereas the gray curve was obtained after imposing 100,000 compression/extension cycles on the membrane. The blue dotted line shows the force prediction on a soapy liquid film on a rigid frame of width $W$ and the black dotted line is obtained by considering a surface minimization with iso-perimetric constraint on its free edges (see Supplementary Material). \textbf{B1.} Soapy liquid catenoid between two parallel circular rings. \textbf{B2.} Two states of the catenoid shape adopted by a wicked membrane attached to two parallel circular rings. \textbf{B3.} Neck radius of the catenoid versus distance between the two rings. The green points represent the experimental observation for a wicked membrane. The blue dotted line represents the soapy liquid solution for the catenoid and the black dotted line shows the solution for a catenoid with iso-perimetric constraint (see Supplementary Material). \textbf{C1.} Soapy bubble. \textbf{C2.} Bubble generated by inflating a wicked membrane at two different inflating stages. \textbf{C3.} Pressure versus radius diagram. Here, the radius of a bubble is defined as $R=(\frac{3}{4\pi}V)^{1/3}$. The blue dotted line represents the theoretical pressure for a soapy bubble being inflated through a tube of radius $R_\mathrm{tube}$ (Laplace's law). The green points correspond to the pressure measurements of the wicked membrane. At $R=R_\mathrm{max}$, the pressure diverges due to the inextensibility of the membrane (the black dotted line illustrates this prediction), making this configuration evocative of the hypotonic swelling of neurons \cite{Morris2001}. Note the significant deviation from Laplace's law resulting from the mixed solid-liquid behavior of the membrane. Scale bars on all the photographs: 1 cm.
}
\label{fig:geometrics}
\end{figure}
Under the constraint of constant liquid film volume and imposed compression $\epsilon$, we minimize the total energy of the system, consisting of the membrane elastic energy $E_\text{el} = \frac{1}{2} B \int \kappa^2 \mathrm ds$ and the surface energy $E_\gamma = 2 \gamma S$, with $\kappa$ and $S$ denoting respectively the local membrane curvature and the exposed surface of the liquid film per unit depth (see details in Supplementary Material). The model reveals $h/L_\text{ec}$ as the relevant parameter governing the behavior of the system, with $L_\text{ec}=\left(B/\gamma\right)^{1/2}$ being the elastocapillary length \cite{Roman2010}.
The limit $h/L_\text{ec} \ll 1$ typically corresponds to the behavior of everyday life soaked fibrous membranes (e.g. wet paper or cloth), that just sag or buckle globally when compressed, irrespective of any surface tension effects (see Fig.~\ref{fig:membrane_reserve_mechanics}C).
Conversely, our experiment is characterized by values of $h/L_\text{ec} \gg 1$ for which the microstructure differs markedly from the previous one: interface energies cannot longer be neglected and the membrane now buckles under the capillary confinement of the interfaces (see Fig.~\ref{fig:membrane_reserve_mechanics}C).
Indeed, in this regime the ratio of the surface energy to the elastic energy scales as $E_\gamma/E_\text{el}\sim\left( h/L_\mathrm{ec} \right)^2 \gg 1$
, i.e. any deformation of the liquid surface introduces a strong energetical penalty, making it clear that in-film wrinkling is a low energy state configuration.
This phenomenon is therefore reminiscent of buckling under rigid confinement, for which wavelengths $\lambda$ also scale linearly with the confinement gap $h$ for a given compression $\epsilon$ \cite{Roman1999}, and this behavior is indeed nicely recovered by our model (see inset of Fig.~\ref{fig:membrane_reserve_mechanics}D).
Somewhat surprisingly, the experimentally measured wavelengths $\lambda$ prove to be insensitive to the compression.
This behavior, not captured by the model, coincides with the emergence of a second, tightly packed, accordion-like phase (see Fig.~\ref{fig:membrane_reserve_mechanics}B).
This second phase, in mechanical equilibrium with the wrinkled phase, has a high membrane storage capacity and actually corresponds to the membrane reserve.
The coexistence between these phases allows to continuously transfer material from one phase to another, and warrants the effectiveness of the wicked membrane as a stretchable material.
\begin{figure}[h!]
\begin{center}
\includegraphics[width=16cm]{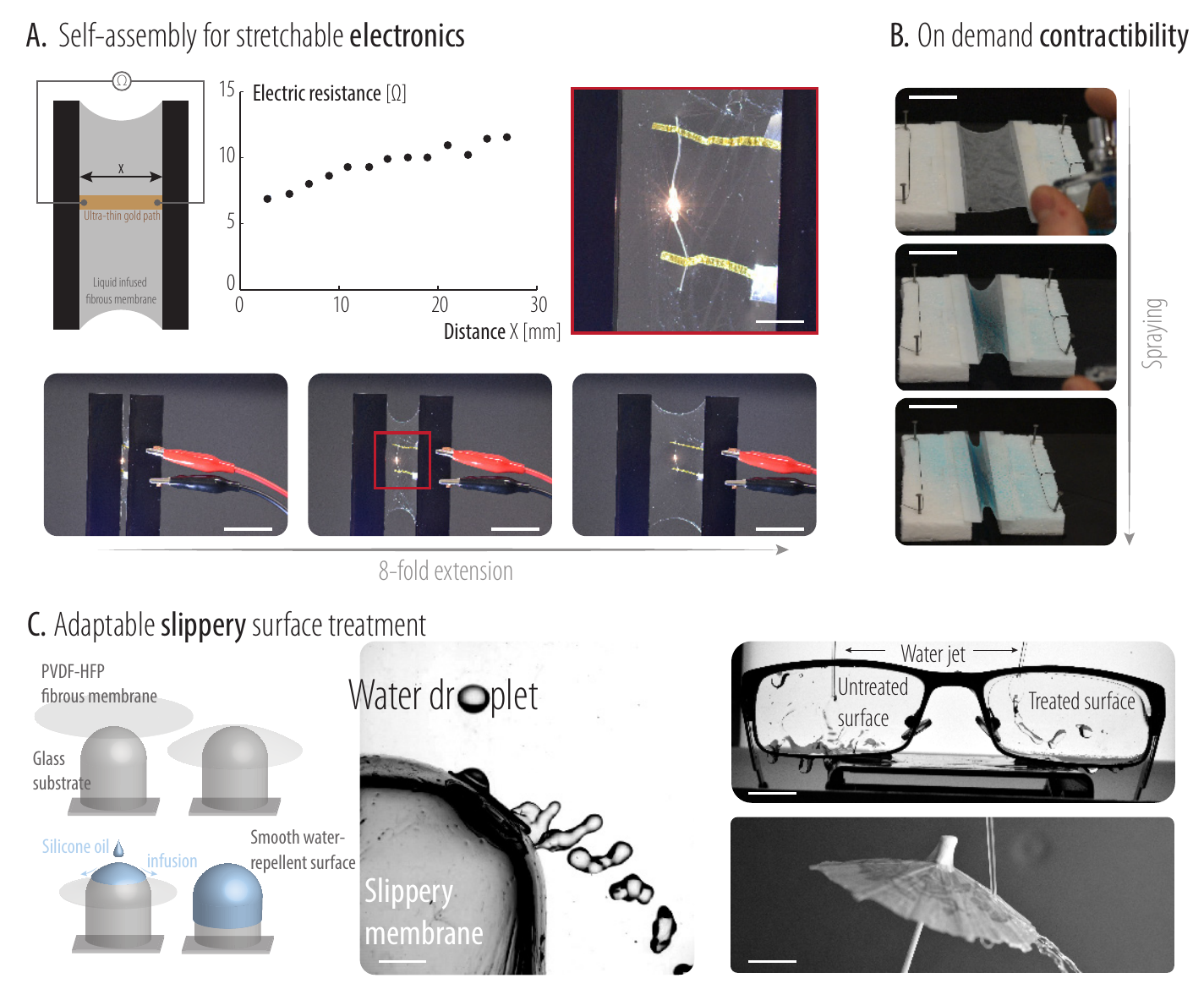}
\end{center}
\caption{\textbf{Stretchable electronics, on-demand contractibilty and adaptable slippery surfaces}. \textbf{A.} A thin gold strip (100 nm thick) is apposed on the wicked membrane (here : PVDF-HFP membrane wicked by v100 silicone oil). Capillary adhesion guarantees the gold strip to remain secured to the membrane. Electricity runs through the gold strip even when membrane reserves are spontaneously generated upon compression of the membrane. Electric resistance versus membrane extension $X$ is presented in the graph and due to local short circuits, this resistance drops as the membrane is contracted. Indeed, as shown on the right photograph (scale bar: 7 mm), a gold strip wrinkles and folds when the membrane is compressed. The sequence shows the illumination of a 1.5V LED through a basic electric circuit throughout an 8-fold extension of the supporting wicked membrane. Scale bar: 3cm. \textbf{B.} Capillary contraction upon wetting. The dry membrane is here attached to two mobile supports (rafts freely gliding on a water bath, first photograph). When ethanol (dyed blue for visualization) is sprayed on the membrane, membrane reserves are immediately generated through the capillary action; the two mobile rafts are pulled towards one and other. Scale bar: 3cm. \textbf{C.} A PVDF-HFP fibrous membrane is apposed to a non-flat surface and is wicked by silicone oil. Capillary adhesion secures the wicked membrane onto the surface and the spontaneous generation of surface reserves allow for a macroscopically smooth surface even on a warped surface (folds are confined in the liquid film). The wicked membrane then acts as a slippery surface treatment with water repellent properties. The chrono-photography (10 ms intervals, scale bar: 7mm) shows a water droplet bouncing off such a treated spherical glass surface. The wicked membrane can adapt to any object such as glasses (scale bar: 1.5 cm, water droplets stick to the untreated surface but roll off the treated glass) or cocktail umbrellas (scale bar: 2cm, the water repelling property is conserved after consecutive opening/closing of this toy umbrella).}
\label{fig:applications}
\end{figure}
The local mechanics just described reflects into global geometries and force response that we explore next in Figure~\ref{fig:geometrics}.
There we subject wicked membranes to three different elementary solicitations, corresponding to stretching of planar-, cylindrically-, and spherically-shaped membranes.
Strikingly enough, the equilibrium shape adopted by the wicked membrane in each configuration strongly resembles that of a liquid film under the same conditions: planar film, catenoid and bubble.
Once again, this behavior is made transparent by realizing that in the limit $h/L_\text{ec} \gg 1$ the energy of the wicked membrane is dominated by its capillary contribution; the equilibrium shapes therefore essentially correspond to minimal surfaces.
Although they differ strongly in longevity, ways of fabrication and internal structure, the wicked membranes and liquid films therefore present interesting similarities.
Upon closer inspection though, the shapes of the membranes appear to differ from their liquid counterparts in some respects.
For example, a planar membrane attached on only two straight edges adopts a stable shape (Fig.~\ref{fig:geometrics}A), whilst a liquid film in the same configuration would merely burst.
To understand this stabilization mechanism for membranes, we have to perceive that some regions of the membrane may undergo stretching up to a point where the membrane reserves are fully exhausted.
Pure stretching deformations represent a far higher energetical cost as compared to bending ones \cite{Audoly2010} and
as a first approximation, this sharp energetical penalty can be seen as an inextensibility constraint.
And indeed, the shapes adopted by the planar configuration can fully be captured with a surface area minimization under isoperimetric constraint (see Supplementary Material).
This mixed liquid-solid behavior allows to stabilize the catenoid shape beyond its classic point of bursting to unveil new equilibria (Fig.~\ref{fig:geometrics}B and Supplementary Material), and is also responsible for significant deviations from Laplace's law in the bubble configuration (Fig.~\ref{fig:geometrics}C).
Such a hybrid mechanical behavior is again reminiscent of the response of cellular membranes, and indeed, whether for lymphocytes, fibroblasts \cite{Guillou2016,Groulx2006} or wicked membranes (Fig.~\ref{fig:geometrics}), the mechanical response switches from liquid-like to solid-like once all the membrane reserves have been flattened out.

The peculiar behavior of our wicked membrane stems from its compound nature: while surface tension allows it to undergo ample shape changes, its solid underlying matrix provides mechanical robustness. Geometrical reorganizations at the microstructural level (reserve self-assembly or unfolding) are key in the mechanics of the wicked membrane, and allow in particular to prevent any significant stretching at the molecular level. This results in a marked resilience of this material to fatigue, as illustrated by the unvarying mechanical response of the membrane after more than 100,000 cycles of 10-fold stretching and compression events (Fig.~\ref{fig:geometrics}A3 and Supplementary Materials).

Moreover, it is worth emphasizing that the mechanical functionalization of a membrane into a highly stretchable membrane only relies on a combination of elasticity, capillarity and geometry. As such, the process is widely universal (see the list of tested polymers in the Supplementary Material), and opens appealing prospects from an engineering point of view. To illustrate but a few applications taking advantage of this mechanical function, we present some proofs of concepts in Figure~\ref{fig:applications}.
First we demonstrate that affixing submicronic thick golden paths to the membrane allows to effortlessly obtain a $\sim$10-fold stretchable conductive material, presenting an unusual reversibility.
Interestingly, due to the formation of folds and membrane reserves, the effective electrical resistance of the material displays a dependence to its stretching state (Fig.~\ref{fig:applications}A).
Second, by controlling the formation of the liquid film (via the surrounding humidity level, or with the instant release of a liquid), on-demand contractility of the wicked membrane is instantaneously obtained, making it straightforward to apply forces, possibly while distorting the membrane (Fig.~\ref{fig:applications}B).
Finally we show that a wicked membrane infused with a lubricant becomes a transparent slippery surface \cite{Okada2014} that can provide an instant non-wetting functionalization to curved, rough, warped and deformable surfaces thanks to its exceptional shape adaptability (Fig.~\ref{fig:applications}C).

The use of membrane reserves to fuel large shape changes is encountered in an extremely wide variety of animal cells, but this strategy has so far not been used to create stretchable synthetic materials. The ease of manufacturing, self-assembly of membrane reserves, universality of the process along with its robustness opens tremendous novel pathways for the design of synthetic stretchable devices (possibly biocompatible), electronics, batteries or actuators.

\putbib[membrane_reserve]
\end{bibunit}

\begin{bibunit}[plain]
\newpage
\newpage

\section*{Materials}
The polymers used to fabricate the fibrous membranes are Poly(vinylidene fluoride-co-hexafluoropropylene) (PVDF-HFP, Solvay), Polyacrylonitrile (PAN, M.W. 150,000, Sigma Aldrich), Polycaprolactone (PCL, M.W. 80,000, Sigma Aldrich) and Polyvinylpyrrolidone (PVP, M.W. 1,300,000, Acros Organic). The solvents are n,n-dimethylformamide (DMF, Carlo Erba Reagents) and ethanol (absolute, Sigma Aldrich). The liquids used to infuse the fibrous membranes are deionized water, glycerol (Sigma Aldrich), ethanol (absolute, Sigma Aldrich) and PDMS v100 oil (Sigma Aldrich). The concentration of polymer of each polymer/solvent solution is 10\% wt. A summary of the constituents is provided in Table \ref{table:polymers}. Surface tensions of the liquids are characterized using a Krüss K6 manual tensiometer (hanging ring). 
Erioglaucine disodium salt (Sigma Aldrich) is used to dye the fibrous membrane (dissolution of the coloring agent in the polymer/solvent solution prior to electrospinning) and the infusing liquids. 1mm wide gold strips are obtained by manually cutting 100 nm thick edible gold leaves (purchased from Alice Delice) using a surgical blade.

\begin{table}[h!]
\begin{center}
\begin{tabular}{  c c c  } 

 \textbf{Polymer} & \textbf{Solvent} & \textbf{Wicking liquid} \\ 
 \hline
 PAN & DMF & water - glycerol - ethanol - silicone oil \\ 
 PVDF-HFP & DMF & ethanol - silicone oil \\ 
 PCL & DMF & ethanol - silicone oil \\
 PVP & ethanol &  silicone oil
\end{tabular}
\caption{Polymers used to fabricate fibrous membranes with their respective solvents and infusing liquids that are used to generate self-assembled surface reserves.}
\label{table:polymers}
\end{center}
\end{table}

\section*{Sample preparation}
The fibrous membranes are obtained using an electrospinning technique \cite{Greiner:2007} using the electrospinning apparatus ES-1A (Electrospinz Ltd.) following these key steps: 
\begin{enumerate}
\item A polymer is dissolved in a solvent (the polymers and corresponding solvents that are used in this work are presented in Table \ref{table:polymers}). 
\item The solution is injected through an electrically charged blunt needle (diameter of the needle 1mm at a rate of 0.02 ml/min, between 10 and 15kV). The outgoing droplet is instantaneously destabilized through the formation of a Taylor cone which is ejected as a liquid rod towards an electrically neutral fixed plane target (distance between the tip of the needle and the target: 17cm). As it travels towards the target, the solvent evaporates from the liquid rod which therefore quickly undergoes a swirling instability randomly deviating it. 
\item The resulting fibrous mat (made of the addition of solid fibers continuously generated) is recovered from the target, which was previously covered with anti-adhesive cooking paper (purchased from Monoprix S.A.) to avoid sticking. 
\end{enumerate}

Once the membrane is attached to the mobile supports (Thorlabs translational elements or laser cut PMMA assembly of 8 translational supports), it is wicked by a wetting liquid (see Table \ref{table:polymers}) using a Terumo 10ml syringe or a spray. Upon compression, the surface reserves are instantaneously formed through the wrinkling and folding of the membrane under the capillary forces.

\section*{Thickness characterization of the infusing liquid film}
\begin{figure}[h] 
\begin{center}
 \includegraphics[width=.9 \textwidth]{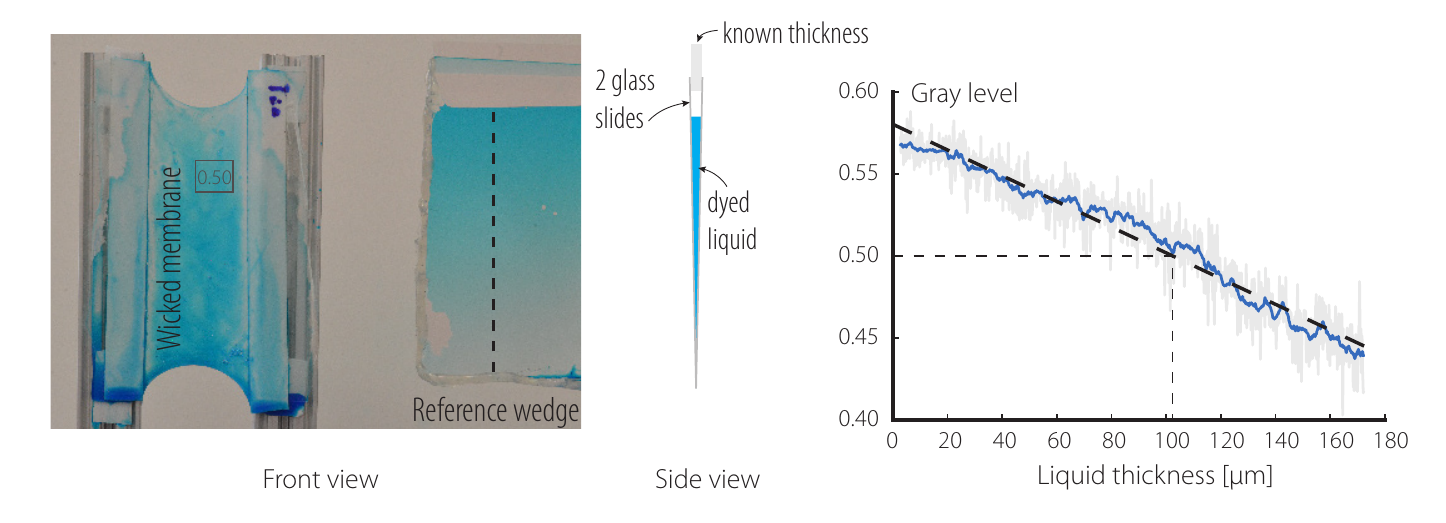}
 \caption{\label{fig:grayscale} Measurement the liquid film thickness of the infused membrane with a {colorimetry} method. A scale wedge (two slightly non parallel glass slides) is used to calibrate the gray level as a function of colored liquid thickness on a photograph. In this case, we show an area on the membrane where the gray level is 0.50, corresponding to a liquid film thickness of  around 100 $\mu$m. }
\end{center} 
\end{figure}

For the study the wrinkling wavelength $\lambda$ as a function of the liquid thickness $h$, a colorimetry tool is used to characterize the liquid film thickness. The membrane is wicked by a dyed liquid (water dyed blue) and a photograph of the infused membrane is taken next to a calibration wedge containing the same dyed liquid with a D810 Nikon camera. Comparing the photograph's local gray value on the membrane and the thickness versus the gray value curve (see figure \ref{fig:grayscale}), we can locally estimate the thickness of the liquid film. Image analysis is performed using the image processing package Fiji.

\section*{Wavelength measurement}
\begin{figure}[h] 
\begin{center}
 \includegraphics[width=.9 \textwidth]{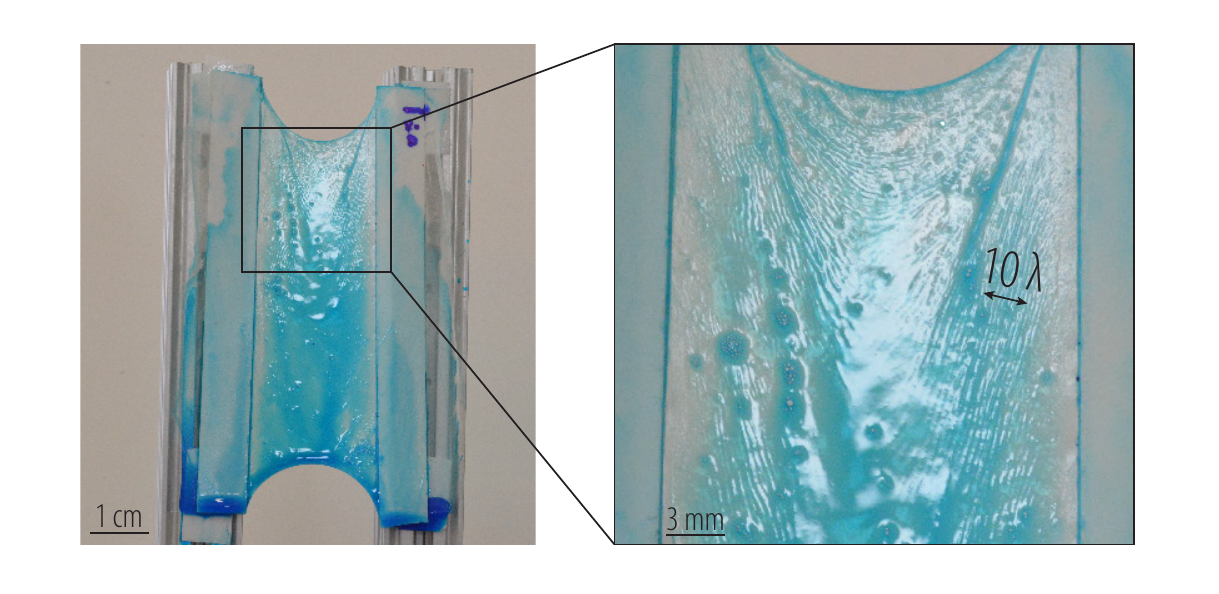}
 \caption{\label{fig:wrinkle}Visualization of the wrinkles on a slightly compressed membrane. }
\end{center} 
\end{figure}

The wrinkling wavelength $\lambda$ is measured when the wicked membrane is slightly compressed. The membrane is illuminated from the side in order to enhance the wrinkles' contrast and a photograph is taken with a D810 Nikon Camera. For each liquid film thickness (different amount of wicked liquid), a set of 4 wavelength measurements is performed. The coloring agent (erioglaucine disodium salt) did not show to change the liquid surface tension significantly. A typical wavelength measurement image is presented in figure \ref{fig:wrinkle}. 

\section*{Force measurement and fatigue characterization}
\begin{figure}[h] 
\begin{center}
 \includegraphics[width=.7 \textwidth]{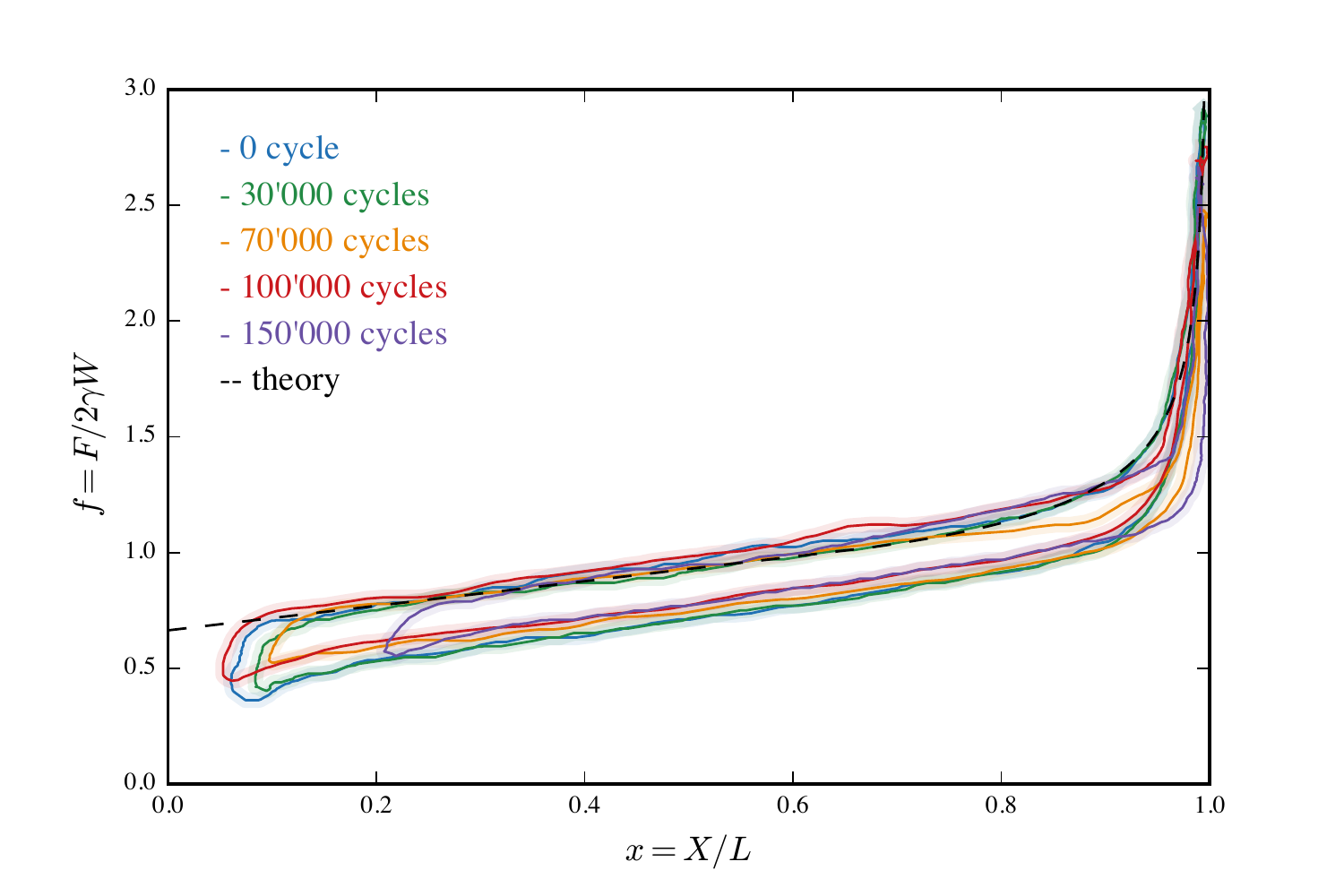}
 \caption{\label{fig:fatigue}Force vs. displacement of the membrane during compression/extension cycle after imposing numerous cycles to it.}
\end{center} 
\end{figure}

The force versus displacement curve of the plane wicked membrane is performed using a cantilever beam method on a silicone oil infused PVDF-HFP fibrous membrane. The cantilever beam's mechanical response was calibrated using calibrated weights (weighed with a Mettler-Toledo MS 0.01 mg precision scale). The membrane is supported by two floating rafts on a water bath to ensure frictionless translational supports. The presented extension/compression force measurement cycles are performed at around 1mm/s but show little sensitivity to displacement speed (the same force vs. displacement curve was obtained for a twice as fast displacement speed). 
\\The fatigue test was performed on a silicone oil infused PVDF-HFP fibrous membrane mounted on a crank rod system and the wicked membrane underwent 3 extension/compression cycles per second. The compression/extension cycle corresponds to an end to end distance X varying from $X_\mathrm{min}=$ 2 mm to $X_\mathrm{max}=$ 3.7 cm. The membrane was re-infused by silicone-oil every 20,000 cycles to avoid drying. Small circular  holes (hundreds of microns in diameter) appeared at around 60,000 cycles, slowly growing up to 150,000 cycles. At this last point, the holes had a significant impact on the mechanical behavior of the wicked membrane (see fig \ref{fig:fatigue}) and the membrane tore off shortly after.

\section*{Catenoid neck radius measurement}
\begin{figure}[h] 
\begin{center}
 \includegraphics[width=1 \textwidth]{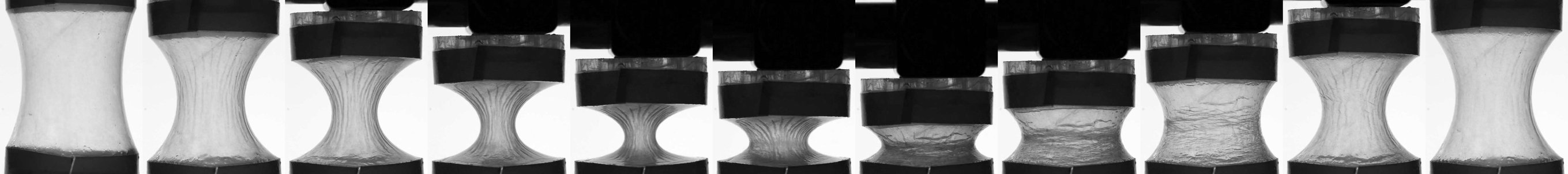}
 \caption{\label{fig:catenoid} Side view of the catenoid-like shape adopted by a wicked membrane attached to the edge of two discs (diameter of the disc: $2R=2.5$ cm) throughout a compression/extension cycle. Note that the this catenoid jumps from a thin to wide state between image 5 and 7. This sudden change in shape is due to the iso-perimetric constraint, responsible for a strong path dependence of the equilibrium state (see Supplementary Material).}
\end{center} 
\end{figure}

To study the equilibrium forms adopted by a wicked membrane in a cylindrical configuration, a PAN membrane is attached on the edge of two laser cut PMMA discs (diameter $2R=2.5$ cm) with 3M double face tape. The membrane is wicked with deionized water and the distance between the two discs is controlled using a Thorlabs 25 mm manual translation stage. Throughout the compression/extension cycle, the shape adopted by the wicked membrane is filmed with a Nikon D810 camera and image post-processing (Python) allows to gather the neck radii of the resulting catenoid for the visited disc distances. Figure \ref{fig:catenoid} shows a sequence of such  compression/extension cycle and makes it clear that the shape adjustments of the wicked membrane are mediated by the self-assembled folds that act as surface reserves.

\section*{Wicked membrane bubble pressure measurement}
To measure the pressure inside an inflated spherical wicked membrane (PAN fibrous membrane wicked with deionized water), visualization of an adjacent ethanol filled tube is used. The bubble is inflated with air using a PHD Ultra Syringe Pumps (Harvard apparatus) at a rate of 6 ml/min. The air-entrance tube is connected to a U-shaped tube partially filled with dyed ethanol with a T-junction. One end of the U-shaped tube is therefore pneumatically linked to the bubble, while the other end is open (at atmospheric pressure). The ifference in height $\Delta h$ of the two ethanol interfaces inside the U-shaped tube indicates the pressure $P$ inside the bubble, knowing its density $\rho=789$ kg/m$^3$ and earth acceleration $g=9.81$ m/s$^2$  ($P=\rho g \Delta h$). The pressure $P$ is normalized by $P_\mathrm{max}=4\gamma/R_\mathrm{tube}$, which is the theoritical maximum pressure for a spherical bubble of surface tension $\gamma$, inflated out of cylindrical tube of radius $R_\mathrm{tube}$ ($R_\mathrm{tube}=$ 4.5 mm in our experiment). It is to be mentioned that in contact with the PAN membrane, the surface tension of deionized water drops from 72 to 53 mN/m. 
\begin{figure}
\begin{center}
 \includegraphics[width=1 \textwidth]{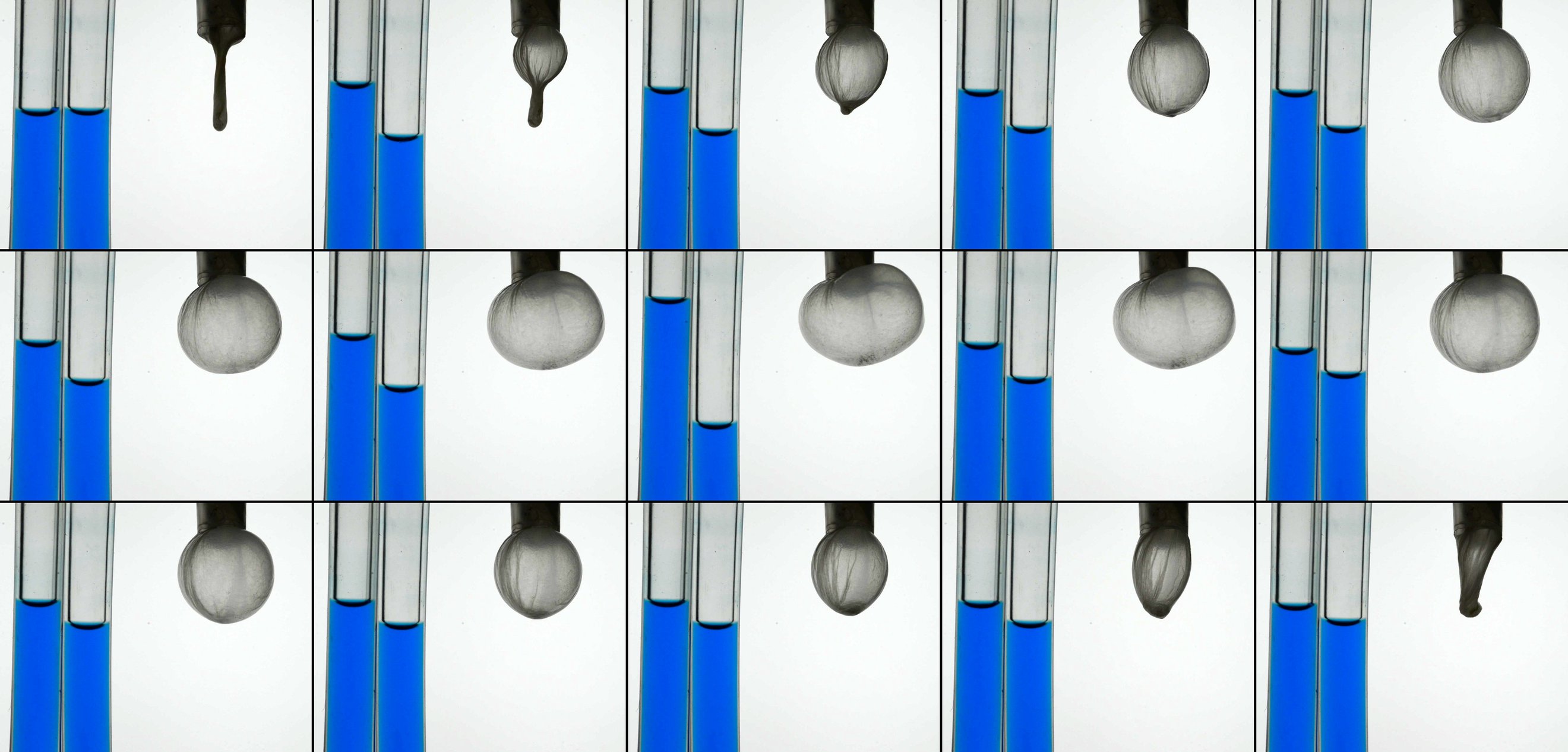}
 \caption{\label{fig:bubble} Side view of the bubble-like wicked membrane inflated with air. The tubes filled with blue dyed ethanol allow to characterize pressure inside the wicked membrane bubble throughout its inflating. The external diameter of the bubble supporting tube is 9 mm. }
\end{center} 
\end{figure}

\newpage \newpage
\section*{Wavelength $\lambda$}
Here, we study the buckling of an elastic beam confined inside a liquid film. Elastic energy minimization would favor high wavelength but for flexible enough beams, capillary interface energy of the liquid film can become non-negligible; the system will find a trade-off wavelength, minimizing the sum of the elastic and capillary energies. 
\begin{figure}[h] 
\begin{center}
 \includegraphics{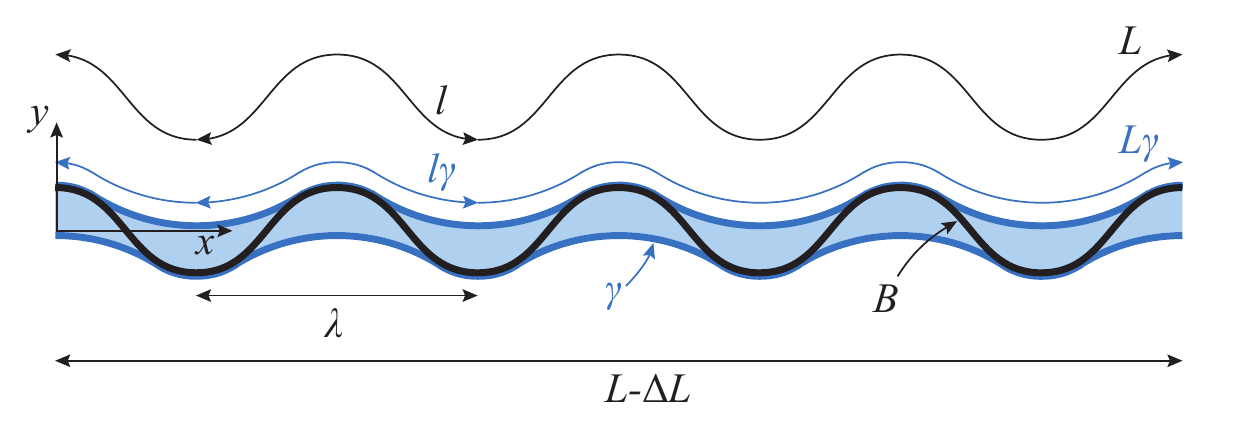}
 \caption{\label{fig:global/view} Wrinkling of a beam of length $L$ under compression ($\Delta L$) in a liquid film of volume $V=h L$. The liquid/air and solid/air interfaces have the same specific interface energy $\gamma$ and the liquid/solid interface is considered to have a specific energy of 0. The beam has a bending resistance $B$ and the texture wavelength is $\lambda$. }
\end{center} 
\end{figure}
The solid beam is described by $y(x)$ whereas the liquid/air + solid/air interface is given by $y_\gamma(x)$. Indeed, since the liquid/air and solid/air interface specific energies are considered to be the same, $y_\gamma(x)$ corresponds to the liquid/air interface where the beam is covered by liquid, but is given by the beam surface $y(x)$ where no liquid covers the beam. 
\subsection*{Energies}
Because the wavelength $\lambda$ is much smaller than the total beam length $L$, boundary conditions at the ends of the beam are not considered. The elastic and interface energies respectively are given by $\Ee$ and $\Eg$ : 
\begin{equation}
\Ee=\frac{1}{2} B \int^{L-\Delta L}_{0}{\kappa^2}(x)\,dx
\label{Ee1}
\end{equation}
\begin{equation}
\Eg= 2 \gamma \int^{L-\Delta L}_{0} \sqrt{1+{{\yg}'}^2(x)}\,dx
\label{Eg1}
\end{equation}
where $\kappa(x)$ represents the beam curvature and $\yg(x)$ represents the bottom liquid/air+solid/air interface. The factor $2$ in eq. (\ref{Eg1}) comes from top/down pseudo-symmetry (the bottom interface has the same length as the top interface).
\subsection*{Volume conservation}
Incompressibility of the liquid requires for the volume $V=h L$ to be conserved: 
\begin{equation}
2 \int^{L-\Delta L}_{0} \left[ \yg(x)- y(x) \right] \, dx = V
\label{h01}
\end{equation}
\subsection*{Inextensibility constraint}
Inextensibility of the beam provides a constraint on its length: 
\begin{equation}
\int^{L-\Delta L}_{0} \sqrt{1+{y'}^2(x)} \,dx = L
\label{L1}
\end{equation}
\subsection*{Refocusing the problem on one wavelength $\lambda$}
$l$ corresponds to the beam length in one wavelength, therefore, it satisfies 
\begin{equation}
\frac{l}{L}= \frac{\lambda}{L (1-\epsilon)}
\end{equation}
with $\epsilon=\frac{\Delta L}{L}$, and equations \ref{Ee1} to \ref{L1} can be rewritten considering only the energies and shapes on one wavelength:
\begin{equation}
\Ee=\frac{1}{2} \, L \frac{1-\epsilon}{\lambda} B \int^{\lambda}_{0}{\kappa^2}(x)\,dx
\label{Ee2}
\end{equation}
\begin{equation}
\Eg= 2 \, L \frac{1-\epsilon}{\lambda}  \gamma \int^{\lambda}_{0} \sqrt{1+\yg'^2(x)}\,dx
\label{Eg2}
\end{equation}
\begin{equation}
h= 2 \frac{1-\epsilon}{\lambda} \int^{\lambda}_{0} \left[ \yg(x)- y(x) \right]\,dx
\label{h02}
\end{equation}
\begin{equation}
\frac{\lambda}{1-\epsilon}= \int^{\lambda}_{0} \sqrt{1+y'^2(x)} \,dx
\label{L2_}
\end{equation}
\\
\subsection*{Normalizing}
Using $\Lec=\sqrt{B/\gamma}$ as unit length and $B/\Lec$ as unit energy, we introduce the following dimensionless quantities:
\begin{subequations}
\label{eq:adim}
\begin{align}
\tilde x = \frac{x}{\Lec} \; ; \quad 
\tilde y = \frac{y}{\Lec} \; ; \quad 
\tilde \yg = \frac{\yg}{\Lec} \; ; \quad \\
\tilde L = \frac{L}{\Lec} \; ; \quad 
\tilde \lambda = \frac{\lambda}{\Lec} \; ; \quad 
\tilde h = \frac{h}{\Lec} \; ; \quad 
\tilde \kappa = \kappa \Lec \; ; \quad \\
\tilde E_e = \frac{E_e \Lec}{B}\; ; \quad 
\tilde E_\gamma = \frac{E_\gamma \Lec}{B} \; ; \quad 
\end{align}
\end{subequations}
Equations (\ref{Ee2}) to (\ref{L2}) can be rewritten as follows:
\begin{equation}
\tilde{\Ee}=\frac{1}{2} \, \tilde{L} \frac{1-\epsilon}{\tilde{\lambda}} \int^{\tilde{\lambda}/2}_{-\tilde{\lambda}/2}{\tilde{\kappa}^2}(\tilde{x})\,d\tilde{x}
\label{Ee2_2}
\end{equation}
\begin{equation}
\tilde{\Eg}= 2 \, \tilde{L} \frac{1-\epsilon}{\tilde{\lambda}} \int^{\tilde{\lambda}/2}_{-\tilde{\lambda}/2} \sqrt{1+\tilde{\yg}'^2(\tilde{x})}\,d\tilde{x}
\label{Eg2_2}
\end{equation}
\begin{equation}
\tilde{h}= 2 \frac{1-\epsilon}{\tilde{\lambda}} \int^{\tilde{\lambda}/2}_{-\tilde{\lambda}/2} \left[ \tilde{\yg}(\tilde{x})- \tilde{y}(\tilde{x}) \right]\,d\tilde{x}
\label{h02_}
\end{equation}
\begin{equation}
\frac{\tilde{\lambda}}{1-\epsilon}= \int^{\tilde{\lambda}/2}_{-\tilde{\lambda}/2} \sqrt{1+\tilde{y}'^2(\tilde{x})} \,d\tilde{x} 
\label{L2}
\end{equation}
\subsection*{Interface $\yg(x)$} \label{subsection*:surface}
\begin{figure}[h] 
\begin{center}
 \includegraphics{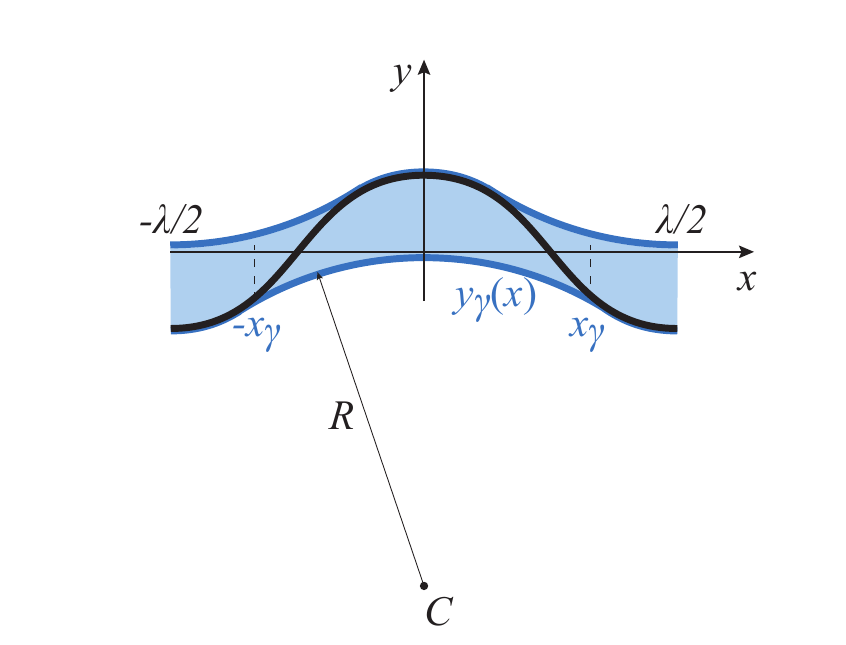}
 \caption{\label{fig:zoom_lambda} Local problem on one wavelength $\lambda$ with new unknowns $\xg$ and $\yg(x)$}
\end{center} 
\end{figure}

The liquid interface and the elastic beam merge at the x-coordinate $\xg$ (see figure \ref{fig:zoom_lambda}). Because $\yg(x)=y(x)$ where no liquid covers the beam, $\xg$ has to satisfy the adapted formulation of equation (\ref{h02_}) (volume conservation): 
\begin{equation}
 2 \frac{1-\epsilon}{{\lambda}} \int^{\xg}_{-\xg} \left[ {y}({x})-{\yg}({x}) \right]\,d{x} = h
\label{eq:x1}
\end{equation}
\noindent We abandoned the "$\sim$" notation for clarity.\\
Since pressure has to be constant inside the liquid film, $\yg(x)$ is defined as a circular arc between $-x_\gamma$ and $x_\gamma$ (of radius $R$). We assume continuity between the liquid interface and the beam at the triple point (no penetration of the beam across the liquid interface) :
\begin{equation}
y_\gamma(x_\gamma) = y (x_\gamma)
\end{equation}
and a tangent connection between the liquid surface and the beam (the liquid wets the beam perfectly):
\begin{equation}
y_\gamma'(x_\gamma) = y '(x_\gamma)
\end{equation}
These two conditions provide an explicit formulation for $y_\gamma(x)$ :
\begin{equation}
\yg(x)=\sqrt{R^2-x^2}-y_\mathrm{c}
\label{eq:yg}
\end{equation}
where $\yc$, the y-coordinate of the center of the circle describing the liquid interface, is given by: 
\begin{equation}
y_\mathrm{c}=y(\xg) + \frac{\xg}{y'(\xg)}
\label{eq:yc}
\end{equation}
and the radius of said circle is expressed as followed:
\begin{equation}
R=\xg\sqrt{1+\frac{1}{y'^2(\xg)}}
\label{eq:R}
\end{equation}
A schematic representation of the liquid surface $\yg(x)$ is presented in figure \ref{fig:zoom_lambda}.
\subsection*{Summary}
The goal is to find the wavelength $\lambda$ which minimizes the total energy (elastic + capillary) given by : 
\begin{equation}
E=\int^{\lambda/2}_{\lambda/2}\frac{1}{2}\kappa^2(x) + 2 \sqrt{1+\yg'^2(x)} dx
\label{eq:Etot}
\end{equation}
where the constant factor $L(1-\epsilon)$ of eqs. (\ref{Ee2_2}) and (\ref{Eg2_2}) is discarded because it does not affect the minimization.
\\ \\The inextensibility constraint on the beam reads:
\begin{equation}
\int^{{\lambda}/2}_{-{\lambda}/2} \sqrt{1+{y}'^2({x})} \,d{x} = \frac{{\lambda}}{1-\epsilon}
\label{L3}
\end{equation}
and the volume conservation is given by:
\begin{equation}
 2 \frac{1-\epsilon}{{\lambda}} \int^{\xg}_{-\xg} \left[ {y}({x})-{\yg}({x}) \right]\,d{x} = h
\label{eq:x1_2}
\end{equation}
The liquid surface $\yg$ is derived in section \textbf{Interface $\yg$}\ref{subsection*:surface}.
\\ \\The unknowns of the problem are $y(x)$, $\lambda$ and $\xg$.
\subsection*{Energy minimization for a sinusoidal buckling pattern}
In order to simplify the problem, we restrict ourselves to sinusoidal buckling patterns of wavelength $\lambda$ for the elastic beam, \textit{i.e.} $y(x)=A\cos \left(\frac{2\pi}{\lambda} x \right)$. These are the energy minimization steps that were followed: 
\begin{enumerate}
\item Choose a guess wavelength $\lambda$ (we define $k=\frac{2\pi}{\lambda}$). The guess beam buckling pattern is given by $y(x)=A\cos(kx)$.
\item The amplitude $A$ of the sinusoid is constrained by the inextensibility constraint. It can be found by numerically solving eq. (\ref{L3}) ($\epsilon$ is known):
\begin{equation}
\int^{{\lambda}/2}_{-{\lambda}/2} \sqrt{1+A^2 k^2 \sin^2\left( k x \right)} \,d{x} = \frac{{\lambda}}{1-\epsilon}
\label{L4}
\end{equation}
\item \label{item:xg} Next, we find $\xg$ that satisfies the volume conversation constraint given by eq. (\ref{eq:x1_2}) ($h$ is known). The integration is performed using a guess value for $\xg$ and using eq. (\ref{eq:yg}):
\begin{equation}
\yg(x)=\sqrt{R^2-x^2}-y_\mathrm{c}
\label{eq:yg_}
\end{equation}
where $y_c$ and $R$ are given by eqs. (\ref{eq:yc}) and (\ref{eq:R}) respectively:
\begin{equation}
y_\mathrm{c}=A\cos\left( k \xg \right) - \frac{ \xg}{A k \sin ( k\xg)}
\label{eq:yc_}
\end{equation}
and the radius of said circle is expressed as followed:
\begin{equation}
R=\xg\sqrt{1+\frac{1}{A^2 k^2 \sin^2 ( k\xg)}}
\label{eq:R_}
\end{equation}
It is to be mentioned that $\yg(x)$ is described by the expression in eq. (\ref{eq:yg_}) for $\abs{x} \le \xg$ but as the beam surface $y(x)$ where $\xg < \abs{x} < \lambda/2 $.
\item We now have to compute the total energy $E$ of the system using equation (\ref{eq:Etot}) where $\yg(x)$ is described as explained in the previous step. The curvature $\kappa$ is defined as $\frac{y''(x)}{(1+y'^2(x))^{3/2}}$. Note that this energy $E$ only depends on the first guess value $\lambda$. 
\item Iterative steps are performed to find the wavelength $\lambda$ that minimizes the total energy of the system. 
\end{enumerate}
\subsection*{Considering a non-zero thickness of a porous beam}
Experimentally, the wrinkling wicked membrane has a non-zero thickness. The thickness of the beam can be taken into account in our model by considering a vertical shift of half a thickness of the liquid interface. Since the membrane is very porous and entirely wicked by the liquid, we consider the that the entire thickness of the beam $t$ adds up to the total liquid volume $V$. Step \ref{item:xg} of the previous section therefore has to be revisited. $y_c$ is shifted as follows: 
\begin{equation}
y_\mathrm{c}=A\cos\left( k \xg \right) - \frac{ \xg}{A k \sin ( k\xg)} - \frac{t}{2}
\label{eq:yc_thickness}
\end{equation}
where $t$ is the thickness of the porous beam. The volume constraint equation given in eq. (\ref{eq:x1_}) rewrites as:
\begin{equation}
 2 \frac{1-\epsilon}{{\lambda}} \int^{\xg}_{-\xg} \left[ {y}(x) - \frac{t}{2}-{\yg}({x}) \right]\,d{x} + t = h
\label{eq:x1_}
\end{equation}
It is to be noted that this formulation is only valid for small deformations (\textit{i.e.} small compressions $\epsilon$).
In this study, we consider the membrane to grow as the liquid film gets thicker, i.e. the membrane thickness $t$ is proportional to the liquid film thickness $h$ wicking it (we use $t=0.8h$). However, we do not consider the bending stiffness per unit depth $B$ to change during this growth. 

\subsection*{Graphical results and discussion}

\begin{figure}[h] 
\begin{center}
 \includegraphics{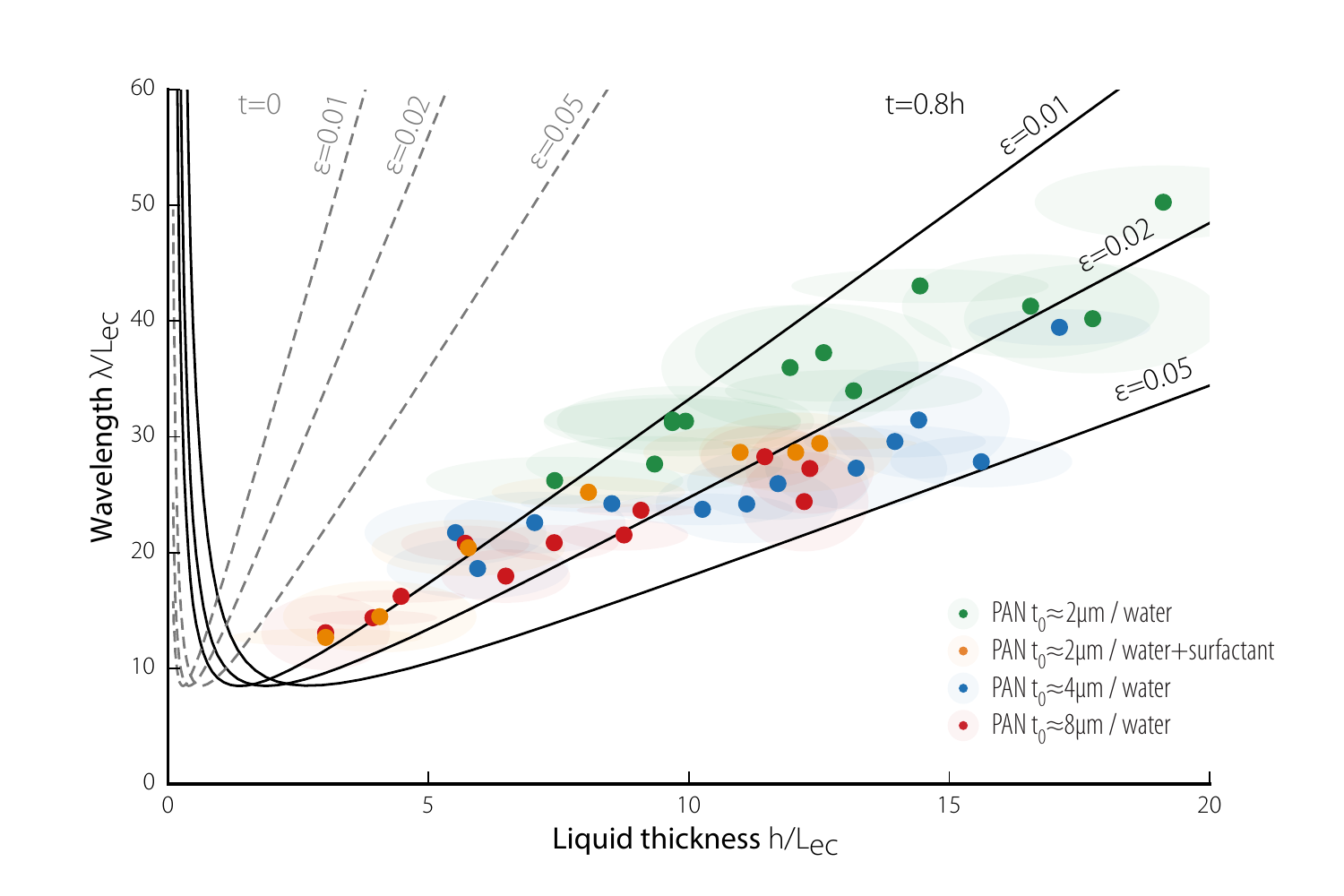}
 \caption{\label{fig:lambda_vs_h} Dimensionless wavelength $\lambda$ versus liquid thickness $h$. The points refer to experimental data for different PAN membrane dry ticknesses and different wicking liquids. The gray solid lines are results of the here presented model for $\epsilon=1$\%, $2$\% and $5$\%. It is considered that the membrane grows as it is infused with liquid. To capture this growth, we choose the membrane thickness $t$ to be proportional to the liquid thickness $h$ ($t=0.8h$). The gray dotted lines represent the same results for a zero-thickness beam. }
\end{center} 
\end{figure}

Figure \ref{fig:lambda_vs_h} presents the dimensionless wavelength $\lambda$ versus the dimensionless liquid thickness $h$ for PAN fibrous membranes of different thicknesses wicked by different liquids. Both $\lambda$ and $h$ are normlized by the elastocapillary length $L_\mathrm{ec}=\sqrt{B/\gamma}$. The surface tension $\gamma$ was measured with the Krüss K6 manual tensiometer (deionized water showed a drop in surface tension when previously put in contact with a PAN membrane, from 72 mN/m to 53 mN/m). To test the dependence of surface tension on the wavelength, the experience was performed with deionized water and a water/soap solution (of measured surface tension $\gamma = $ 30 mN/m). The membrane bending rigidity per unit depth $B$ being low, it could not be measured experimentally. Therefore, it was roughly estimated as $B=\alpha \frac{t_0}{a} E a^3$ where $t_0$ is the membrane dry thickness, $a$ the typical radius of the fiber composing the membrane ($a=500$ $\mu$m) and $E$ is the PAN Young's modulus ($E\simeq 30$ GPa). Finally, $\alpha$ is a dimensionless parameter to account for the membrane porosity (here adjusted to the experiments using $\alpha=2\cdot10^{-4}$). 

\newpage \newpage
\section*{Plane wicked membrane}
Here we consider a wicked fibrous membrane of width $W$ attached to two parallel rigid straight supports at an initial rest distance $L$ from one and other. Upon compression (varying the distance $X$ between the two supports) the membrane remains under tension due to the liquid surface tension and stores the excess membrane inside wrinkles and folds which can afterwards be recruited when extended. The force necessary to keep the two poles at a distance $X$ is here analyzed depending on $W$, $L$ and the surface tension $\gamma$.
\\
\\
Unlike a soapy liquid film on a rigid frame, our membrane is only attached on two edges and has two free edges. It is to be mentioned that a soapy liquid film would break off immediately if it was only ``attached'' on two of its edges. In order to minimize its interface energy with air, the wicked membrane adapts its shape and this minimization with isoperimetric constraint spontaneously leads to circular arcs of arc length $L$ on its free edges \cite{Berest:1998}. Indeed,
the the membrane can wrinkle and fold but cannot be stretched.

\subsection*{Early compression: $\frac{2}{\pi}L<X<L$}
\begin{figure}[h]
\begin{center}
 \includegraphics{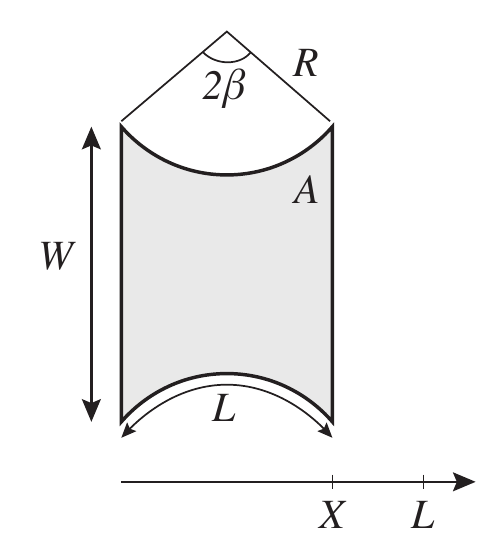}
 \caption{\label{fig:membrane1} Plane wicked membrane during early compression. The surface minimization with iso-perimetic constraint on the free edges of the wicked membrane is responsible for the circular shape adopted by these free edges. }
\end{center}
\end{figure}

At early compression, assuming the free edge describes a circular arc of length $L$ and that the two supports are at a distance $X$, we include two new variables $R$ and $\beta$, respectively the radius and angle span of the circular arc which are presented in figure \ref{fig:membrane1}.
$R$ and $\beta$ are related to $L$ and $X$ through two equations:
\begin{equation}
L=2R\beta
\end{equation}
\begin{equation}
X=2R\sin \beta
\end{equation}
The surface area of the liquid infused membrane then writes:
\begin{equation}
S=WX - 2R^2 \beta + 2RX\cos \beta
\end{equation}
And the normalized force ($f=F/2\gamma W$) that an operator has to apply on the rigid edges to ensure a distance $X$ is given by the derivative of the energy with respect to $X$:
\begin{equation}
f=\frac{1}{2\gamma W} \frac{\partial (2\gamma S)}{\partial X}
\label{eq:force}
\end{equation}
which can be solved numerically.
\subsection*{Advanced compression: $0<X<\frac{2}{\pi}L$}
\begin{figure}[h!] 
\begin{center}
 \includegraphics{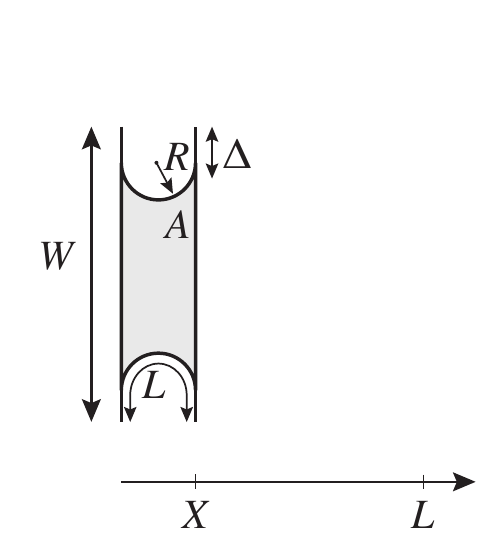}
 \caption{\label{fig:membrane2} Plane liquid infused membrane at a strong compression. }
\end{center}
\end{figure}

\begin{figure}[h]
\begin{center}
 \includegraphics{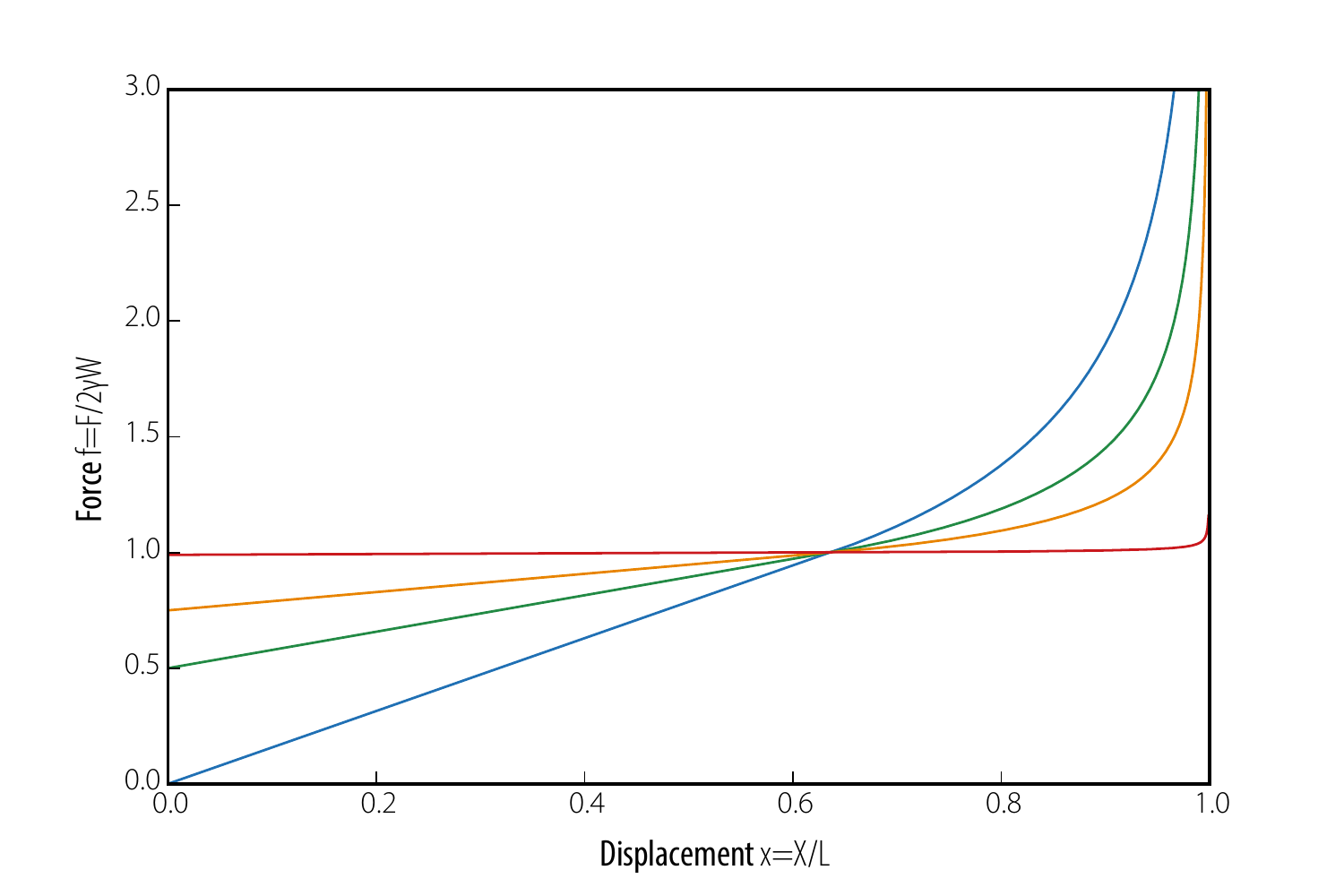}
 \caption{\label{fig:f_vs_x} Theoretical dimensionless force versus displacement curve of the wicked membrane for $L/W=1, \,0.5, \, 0.25, \, 0.01$ for the blue, green, orange and red curve respectively.}
\end{center}
\end{figure}

At a more advanced compression, when $\beta$ reaches $\pi/2$ ({\em i.e.} when $X$ becomes smaller than $2L/\pi$), the geometry adopted by the liquid infused membrane changes and a sketch of it is presented in figure \ref{fig:membrane2}. The radius $R$ of the circular arc remains of interest and a new variable $\Delta$ is considered, it represents the length on which the membrane is sticked to the support.
$R$ and $\Delta$ are given by the 2 relations:
\begin{equation}
L=2\Delta + \pi R
\end{equation}
\begin{equation}
X=2 R
\end{equation}
and again, the surface is calculated:
\begin{equation}
S=WX - 2X\Delta - \pi R^2
\end{equation}
Finally, injecting the three previous relations in the force equation, eq. (\ref{eq:force}), leads to an explicit expression for the force:
\begin{equation}
f = 1-\frac{L}{W} \left (1- \frac{\pi x}{2} \right)
\end{equation}
where x=X/L.
The theoretical dimensionless force vs. displacement curves are given in figure \ref{fig:f_vs_x} for 4 different rest distance to width ratios L/W.

\newpage \newpage
\section*{Catenoid formed by the wicked membrane}
\begin{figure}[h] 
\begin{center}
 \includegraphics{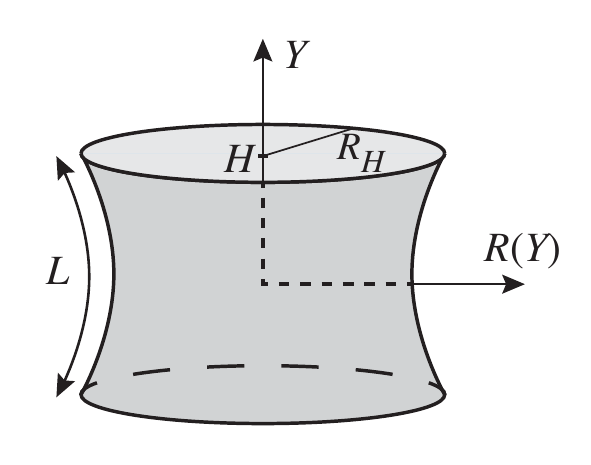}
 \caption{\label{fig:liquid_solid} Sketch of the catenoid-like shape generated by a wicked membrane with self-assembled surface reserves attached to two parallel rings.}
\end{center} 
\end{figure}
Here, we consider a cylindrical geometry, reminiscent of the archetypical liquid soap catenoid : the wicked membrane is attached to two rigid parallel rings of radius $R_H$ at an initial distance $L$. 
Since the membrane is infused with a liquid, the output shape will minimize the total interfacial energy of the system, \textit{i.e.} $2\gamma S$ where $\gamma$ is the liquid-vapor specific energy of the liquid, and $S$ the surface area of the catenoid. As surface tension does not vary, the system seeks to minimize its surface area S given by: 
\begin{equation}
S =  \int_{-H}^{H} 2 \pi R \sqrt{1+R'^2} \, \mathrm{d}Y
\label{eq:surface}
\end{equation}
Moreover, since the fibers of the membrane are inextensible (they can wrinkle inside the liquid film, but cannot be stretched) an inextensibility constraint is to be applied for the outer length of a given fiber from $Y=-H$ to $Y=H$, this length is given by \cite{Berest:1998}:
\begin{equation}
\int_{-H}^{H} \sqrt{1+R'^2} \, \mathrm{d}Y \le L
\label{eq:constraint_length}
\end{equation}
where $L$ is the rest length between the two rings (corresponding to the distance between the two rings when the membrane is perfectly cylindrical). 
The energy of the system can then be re-written as :
\begin{equation} 
\mathcal{V}=  \int_{-H}^{H}  2 \pi (R-\mu) \sqrt{1+R'^2} \, \mathrm{d}Y \ =  2\pi \int_{-H}^{H}  \mathcal{L} (R, R')\, \mathrm{d}Y
\label{eq:V}
\end{equation}
Where $2\pi \mu \ge 0$ is the Lagrange multiplier corresponding to the constraint given in eq. (\ref{eq:constraint_length}). The inextensibibilty constraint is not active when $\mu<0$. Since $ \mathcal{L} (R, R')$ does not depend explicitly on $Y$, minimizing $\mathcal V$ can be done by solving :
\begin{equation} 
\mathcal{H} = \frac{\partial \mathcal{L}}{\partial R'} R' - \mathcal{L} = c
\label{eq:H}
\end{equation}
Where c is a constant to be determined. Deriving the equation leads to:
\begin{equation} 
(R-\mu) \sqrt{1+R'^2} - R' \frac{(R-\mu)R'^2}{\sqrt{1+R'^2}}=c
\label{}
\end{equation}
Which, after solving, leads to :
\begin{equation} 
R(Y) = c \cosh\left(\frac{Y}{c}\right) - \mu
\label{eq:R_Y}
\end{equation}
With two implicit equations for $c$ and $\mu$ :
\begin{equation} 
c \sinh \left(\frac{H}{c}\right) = L
\label{eq:L0}
\end{equation}
\begin{equation} 
c \cosh \left(\frac{H}{c}\right) - \mu = R_H
\label{eq:c_mu}
\end{equation}
which have no explicit solutions but can be solved numerically. 
\\ \\
When $\mu < 0$ the inextensibility constraint becomes inactive and as for the usual liquid catenoid, only the surface area given in eq. (\ref{eq:surface}) has to be minimized. 
\\
The catenoids can be compared by analyzing their neck radii $R(Y=0)$ as a function of the distance between the two rings for given rest lengths $L$ between the rings (distance between the rings for which the membrane is perfectly cylindrical). 
\begin{figure}[h] 
\begin{center}
 \includegraphics{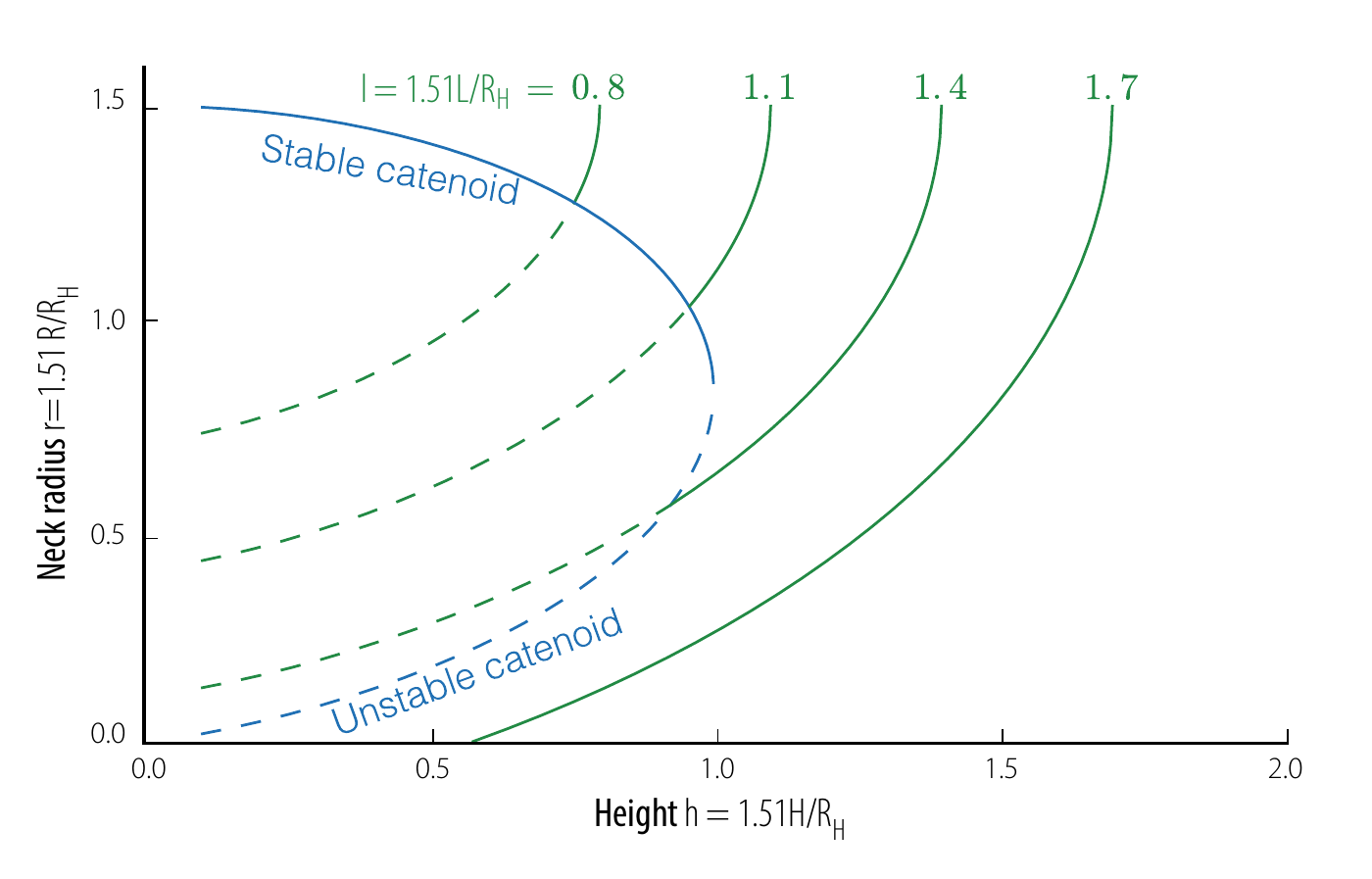}
 \caption{\label{fig:graph_catenoid} \textbf{Dimensionless neck radius $r$ versus dimensionless distance between the two rings $h$} for different length constraints $l$. The length $R_h/1.5090$ is commonly used to normalize lengths in catenoid-related problems.     It is to be noted that the inextensibibilty constraint allows for catenoid geometries to exist in a region where purely liquid catenoids do not exist ({\em i.e.} where $h>1$). Solid green lines represent solutions of eq. (\ref{eq:R_Y}) (the neck radius refers to $R(Y=0))$ for different isoperimetric constraint (distance between the two rings when the membrane is straight $L$) where $\mu \ge 0$, whereas the dotted green lines represent the solution of the same equations but with $\mu<0$ {\em i.e.} where the constraint is inactive. The solid and dotted blue line respectively represent the stable and unstable purely liquid catenoid. }
\end{center} 
\end{figure}
It is to be mentioned that a wicked membrane can display two stable catenoid shapes for given parameters. For example, the curve corresponding to $l=1.4$ shows that for a given height $h$ slightly below 1.0, the neck radius can be that of a pure soapy liquid catenoid (blue curve), or that of a catenoid with isoperimetric constraint ($l=1.4$) corresponding to the green curve. This bistability results in a strong path-dependence for the adopted shape of the catenoid. Indeed, when this catenoid comes from $h >1$, it follows the green curve when bringing the two rings closer ({\em i.e.} $\mu>0$, the isoperimetric constraint is active). When it crosses the unstable liquid catenoid solution (dotted blue line), $\mu$ changes sign and becomes negative, thus making the isoperimetric constraint inactive. The catenoid then jumps to the stable liquid catenoid solution (solid blue curve). On its way back (increasing $h$) it continues traveling only on the blue curve until it crosses the $h=1$ axis, in which case it will jump back to the isoperimetric constrained solution (green curve). 

\newpage
\putbib[membrane_reserve]
\end{bibunit}

\end{document}